
%
%
\magnification=1200

\def\today{\ifcase\month\or
  January\or February\or March\or April\or May\or June\or
  July\or August\or September\or October\or November\or December\fi
  \space\number\day, \number\year}
\font\twelverm=cmr12  \font\twelvebf=cmbx12

\def\bigtype{\let\rm=\twelverm \let\bf =\twelvebf \rm}

\def\hsingle{\baselineskip=18truept}

\overfullrule=0pt
\vsize=8.8truein
\hoffset=.2truein
\hsize=6.1truein
\hsingle

\parskip=5pt plus 2pt
\parindent=20truept

\line{Submitted to {\it Journal de Physique II} (December, 1993)\hfill}
\vskip 1cm
\centerline{\bigtype{\bf
Capillary Rise in Tubes with sharp Grooves
}}

\tenrm
\bigskip\bigskip
\centerline{Lei-Han Tang}
\medskip
\centerline{Institut f\"ur Theoretische Physik,
Universit\"at zu K\"oln}
\centerline{Z\"ulpicher Str. 77, D-50937 K\"oln, Germany}
\bigskip
\centerline{Yu Tang}
\medskip
\centerline{Department of Biophysics, Shanghai Medical University}
\centerline{Shanghai 200032, People's Republic of China}
\bigskip
\vskip .5truecm
{\narrower\smallskip\noindent
{\bf Abstract:}
Liquid in grooved capillaries, made by e.g. inserting a plate in a
cylindrical tube, exhibits unusual spreading and flow properties.
One example is capillary rise, where a long, upward tongue on top of
the usual meniscus has been observed along the groove.
We attribute the underlying mechanism to a thermodynamic instability
against spreading for a (partial or complete wetting) liquid in a sharp
groove whose opening angle $\alpha$ is less than a critical
value $\alpha_c=\pi-2\theta$.
The equilibrium shape of the tongue is determined analytically.
The dynamics of liquid rising is studied in the viscous regime.
When the diameter of the tube is smaller than the
capillary length, the center part of the meniscus rises with time $t$
following a $t^{1/2}$-law,  while the tongue is truncated at a height which
grows following a $t^{1/3}$-law.
Sharp groove also facilitates release of gas bubbles trapped
inside a capillary under the action of gravity.
\smallskip}
\noindent{\it PACS numbers:}
68.10.Gw, 82.65.-i, 87.45.Ft
\bigskip
\vfil\eject
\line{\bf 1. Introduction\hfil}
\medskip

It is well-known that capillary rise in a thin tube
is inversely proportional to the diameter of the tube[1].
This law holds for partial as well as complete wetting.
In this paper we show that,
when a sharp, vertical groove of sufficiently small opening angle
is constructed in the tube, a liquid tongue of macroscopic thickness
appears inside the groove. The tongue extends to arbitrarily high altitudes,
independent of the diameter of the tube.

Our work started with the need to fill thin glass
tubes with certain kind of ionic solution, to be used as microelectrodes
in physiological experiments.
A practical problem is to avoid trapping of gas bubbles in the tube.
When tubes with a circular cross-section are used [see Fig. 1(a)],
it is very difficult to get rid of bubbles particularly in the
sharp tip region. A simple solution was found by
inserting a plate inside the raw tube before stretching it into
the final shape using standard glass-making technique
[see Fig. 1(b)]. Tubes with cross-sectional shapes shown in
Figs. 1(c)-1(e) were found to be equally good for this purpose[2].

The tubes which do not trap gas bubbles have a common feature
that there are sharp grooves inside. When such a tube is held
vertically, liquid on top of a trapped bubble develops tongues which
extend down along these grooves.
As a result, trapped bubbles rise easily to the top under gravity.
The spontaneous flow of liquid along the grooves suggests that
there is no free energy barrier for spreading.
Indeed, when these tubes (about 1 mm in diameter and initially empty)
are placed vertically in contact with the solution,
we discover that, on top of the normal capillary rise,
the liquid creeps up along the grooves to very high altitudes.

The new physics introduced by the groove geometry is
the increased tendency for the liquid to spread.
The bottom part of the grooves which appear in the
constructions shown in Figs. 1(b)-(e) can be approximated
by a wedge of opening angle $\alpha$.
As we shall show below, within the classical thermodynamic theory,
one can identify a critical angle
$$\alpha_c=\pi-2\theta,\eqno(1)$$
where $\theta$ is the equilibrium contact angle.
For $\alpha<\alpha_c$, spreading occurs even when the liquid
does not wet the wall completely.
Depending on the initial state of the liquid, gravity may
either enhance or act against spreading.
In this paper we focus on capillary rise against gravity.
In this case there is an equilibrium liquid-air interface profile
which can be determined analytically.
The dynamics of the rising process will also be considered.

The paper is organized as follows. In section 2 we discuss
the static aspects of the problem. Section 3 contains an analysis
on the dynamics of the rising process in the viscous regime.
Discussion and conclusions are presented in Sec. 4.

\bigskip
\line{\bf 2. The Equilibrium Problem\hfil}
\medskip
\line{\it 2.1. Critical angle for spreading\hfil}
\medskip

In the absence of gravity,
the instability of a liquid drop against spreading in a sharp groove
with an opening angle $\alpha<\alpha_c$ can be seen as follows.
Consider a (fictitious) liquid column
of length $L$ that fills the bottom part of the groove.
The cross-section of the column is taken to be an isosceles triangle
with the two equal sides on the wall and a perpendicular of length $d$.
For $L\gg d$, the total surface free energy of the liquid is given by
$$F=2dL\gamma\tan(\alpha/2)+2dL(\gamma_{SL}-\gamma_{SG})\sec(\alpha/2),
\eqno(2)$$
where $\gamma$, $\gamma_{SG}$, and $\gamma_{SL}$ are the
surface tensions of liquid and gas, solid and gas, and
solid and liquid, respectively.
Using the Young's formula
$$\gamma \cos\theta=\gamma_{SG}-\gamma_{SL},\eqno(3)$$
Eq. (2) may be rewritten as
$$F=2dL\gamma [\sin(\alpha/2)-\cos\theta]/\cos(\alpha/2).
\eqno(4)$$
Equation (4) shows that, for $\sin(\alpha/2)<\cos\theta$
or $\alpha<\alpha_c=\pi-2\theta$
and a fixed volume $V=d^2L\tan(\alpha/2)$,
the free energy of the column
decreases without bound as $L\rightarrow\infty$.
We are thus led to the conclusion that
any liquid drop of finite extent is unstable against
spreading for $\alpha<\alpha_c$.

\bigskip
\line{\it 2.2. Equilibrium shape of the tongue\hfil}
\medskip

When the groove is placed vertically, spreading is countered
by gravity. In a cylindrical tube, the macroscopic rise $z$ of the liquid
is limited to a finite value even in the case of complete wetting.
In contrast, for tubes with sharp grooves,
the finite free energy gain per unit length in the groove
[see Eq. (4)] is able to overcome the gravitational energy $\rho gz$
per unit volume, provided the product $dz$ is sufficiently small.
Thus a thin tongue of the liquid can climb up to an arbitrary height
along the bottom of the groove. The width $d$ of the tongue, however,
should be inversely proportional to the height.
Note that the tongue is not a precursor film.
It is completely describable
within the classical thermodynamic framework.

To determine the equilibrium profile of the liquid-air interface
$z(x,y)$, one has to solve the equation[1]
$$\rho g z=2\gamma H,\eqno(5)$$
subjected to the boundary condition that the contact angle of the
liquid at the wall is $\theta$.
Here $\rho$ is the mass density of the liquid (or more precisely,
the difference between the mass density of the liquid and that of the gas),
and $H$ is the mean curvature of the interface.
To obtain a complete solution to (5) for the geometries illustrated
in Fig. 1, one has to resort to numerical means.
Here we are mainly interested in the profile of the tongues.
The problem can then be simplified by considering liquid rise in an infinite
wedge formed by two planar vertical walls at an angle $\alpha$, as
shown in Fig. 2(a).

Anticipating a slow variation of the cross-sectional shape of the
tongue with the height at high altitudes, one may, in a first approximation,
ignore the interface curvature in the vertical direction.
According to Eq. (5), the horizontal cross-section of the tongue
should then take the form of a dart with a circular base, as illustrated
by the closed area $OCBDO$ in Fig. 2(b). The radius $R$ of the arc is
given by $$R=a^2/(2z),\eqno(6)$$
where $a=[2\gamma/(\rho g)]^{1/2}$ is the capillary length.
The angle between the tangent of the circle and the
wall at the point of intersection is $\theta$.
{} From the geometry one obtains the following equation relating the
distance $d=OB$ and the radius $R$ of the circle,
$$d=R\Bigl({\cos\theta \over\sin{1\over 2}\alpha}-1\Bigr).\eqno(7)$$
Combining (7) with (6), we see that the width of the tongue
is inversely proportional to the height when the opening angle $\alpha$
is less than $\alpha_c$.

It turns out that the leading order correction to (7) can also be
calculated analytically when the ratio $R/z={1\over 2}(a/z)^2$ is
small. For this purpose let us introduce polar coordinates $(r,\phi)$
in the horizontal $xy$-plane centered at $A$,
$$x=X-r\cos\phi,\qquad y=r\sin\phi,\eqno(8)$$
where $X=d+R$ is the distance between $O$ and $A$.
The liquid-gas interface position is specified, for small $R/z$, by
$$r(\phi,z)=R\Bigl[1+\Bigl({R\over z}\Bigr)^2\epsilon(\phi)+...\Bigr].
\eqno(9)$$
The mean curvature $H$ can now be expressed in $\epsilon$.
After some algebra the final result is given by
$$H={1\over 2R}\Bigl[1-\Bigl({R\over z}\Bigr)^2
\Bigl({d^2\epsilon\over d\phi^2}+\epsilon+1-\chi^2+
{3\over 2}(1-\chi\cos\phi)^2\Bigr)+...\Bigr],\eqno(10)$$
where $\chi=X/R=\cos\theta/\sin(\alpha/2)$. Inserting (10) into (5)
and using (6), we obtain, to leading order in $(R/z)^2$,
$${d^2\epsilon\over d\phi^2}+\epsilon+1-\chi^2+
{3\over 2}(1-\chi\cos\phi)^2=0.\eqno(11)$$
Equation (11) has a general solution which respects the
$\phi\rightarrow -\phi$ symmetry,
$$\epsilon(\phi)=C\cos\phi-{5\over 2}+{3\over 2}\chi\phi\sin\phi
+{1\over 2}\chi^2\cos^2\phi.\eqno(12)$$

The constant $C$ in (12) is to be determined by the contact angle $\theta$
at the wall. From the geometry we see that the polar angle $\phi_1$ ($>0$) at
which curve (9) intersects the wall satisfies
$$r\sin(\phi_1+{\alpha\over 2})=X\sin{\alpha\over 2}.\eqno(13)$$
Setting the angle between the interface and the wall (in three dimensions)
to $\theta$, we obtain,
$$\epsilon\cos\theta+{d\epsilon\over d\phi}\cos(\phi_1+{1\over 2}\alpha)
+{1\over 2}\cos\theta(1-\chi\cos\phi_1)^2
=0.\eqno(14)$$
Combining Eqs. (12)-(14) gives
$$C={\cos\theta\over 2\sin{1\over 2}\alpha}
\Bigl[4+3\sin^2\theta
-{3\over 2}(\pi-2\theta-\alpha+\sin 2\theta)
\cot{\alpha\over 2}
\Bigr].\eqno(15)$$
Figure 3(a) shows cross-sectional profiles of the interface calculated
using (12) (solid lines) for the case
$\alpha={1\over 3}\pi$ and $\theta={1\over 12}\pi$.
The heights of the cross-sections are at
$z/a=1.5$, 2, 2.5 and 3.
The convergence to the circular profiles (shown by the dashed lines) is
very fast as $z/a$ increases.
Figure 3(b) shows the vertical profiles of the interface for the
same set of parameter values. Solid line gives the contact line of
the liquid with the wall, while the dashed line is the profile
on the bisector of the wedge.

\bigskip
\line{\bf 3. Dynamics\hfil}
\medskip

When a thin tube with sharp grooves is made in contact with a liquid,
the build-up of the equilibrium height consists of three stages:
(i) an initial ``rush'' into the tube, which typically takes
less than $10^{-2}$sec[3]. The flow in this period can be
quite turbulent. (ii) viscous rising in the center part of the tube
towards equilibrium. (iii) development of tongues.
In the following we only consider stages (ii) and (iii).
Here the dynamics is controlled by the viscous flow
of the liquid which limits transport of matter needed to reach
final equilibrium. For simplicity we assume that processes (ii) and (iii)
are separated in time, though in practice there may well be an overlap.
Our considerations follow closely previous work on wetting dynamics
as reviewed by de Gennes[4] and by Leger and Joanny[5].

\bigskip
\line{\it 3.1. Viscous rise in a circular tube\hfil}
\medskip

Let us first consider rising in a thin circular tube without
grooves. The hydrodynamic equation governing the viscous flow
of the liquid in the tube is given by[6]
$$-\nabla ({p\over \rho}+gz)+\nu\nabla^2{\bf v}=0,
\eqno(16)$$
where $\nu$ is the viscosity coefficient of the liquid.
The inertia term has been ignored. As usual, no-slip boundary
condition on the wall is imposed.

The driving force for the flow is the unbalanced pressure drop
across the liquid-gas interface at height $h$,
$$\delta p=\rho g h-4\gamma\cos\theta/D=\rho g (h-h_{\rm eq}),
\eqno(17)$$
where $D$ is the diameter of the tube and $h_{\rm eq}=2a^2\cos\theta/D$ is
the equilibrium height of the interface[7].
The contact angle $\theta$ has a weak dependence on velocity
which we ignore here[4,5].
In the following we make the further approximation that
{\bf v} has only a vertical component.
For an incompressible fluid, this implies that
$v_z$ depends only on the horizontal coordinates $(x,y)$.
In this case Eq. (16) reduces to the Laplace equation
$$(\partial^2_x+\partial^2_y)v_z=\delta p/(\nu\rho h).\eqno(18)$$
The solution to (18) is given by
$$v_z={1\over 4}[\delta p/(\nu\rho h)][r^2-(D/2)^2],\eqno(19)$$
where $r$ is the distance to the center of the tube.

{} From Eq. (19) we obtain the flow rate per unit area
$${1\over \pi (D/2)^2}\int dx dy\ v_z
=-{1\over 8}{\delta p\over\nu\rho h}\bigl({D\over 2}\bigr)^2.\eqno(20)$$
Identifying (20) with the velocity of the interface and using (17),
we obtain,
$${dh\over dt}={g\over 8\nu}\bigl({D\over 2}\bigr)^2{h_{\rm eq}-h\over h}.
\eqno(21)$$
This equation can be easily integrated to give
$$-{h\over h_{\rm eq}}-\ln\bigl(1-{h\over h_{\rm eq}}\bigr)
={1\over 8\cos\theta}\bigl({D\over 2a}\bigr)^3{t\over t_0},\eqno(22)$$
where $t_0=\nu/(ga)$ is a characteristic relaxation time of the liquid,
typically less than $10^{-2}$ sec.
For $h/h_{\rm eq}\ll 1$ or $t\ll t_0(2a/D)^3$
the rising is diffusive,
$$h={h_{\rm eq}\over 2\cos^{1/2}\theta}
\bigl({D\over 2a}\bigr)^{3/2}\bigl({t\over t_0}\bigr)^{1/2}.\eqno(23)$$
It crosses over to an exponential decay
$$h=h_{\rm eq}\Bigl[1-\exp\Bigl\{-1-
{1\over 8\cos\theta}\bigl({D\over 2a}\bigr)^3{t\over t_0}\Bigr\}\Bigr]
\eqno(24)$$
at $t\simeq  t_0(2a/D)^3$.

The important result of the above calculation is that the typical
rising time of the liquid is inversely
proportional to the third power of the diameter of the tube.
Another interesting phenomenon is that, although the equilibrium
rise $h_{\rm eq}$ is higher when the tube is thinner,
the time it takes to reach a given height $h\ll h_{\rm eq}$
is actually longer, $t(h)\simeq t_0 h^2/(aD)$.
The diffusive regime exists only when the diameter is significantly
smaller than the capillary length.
The approximation that ${\bf v}$ has only a vertical component
is not correct close to the top of the liquid column or at the
root of tongues.
However, we expect that the qualitative behavior of viscous rising
is captured by (22).

\bigskip
\line{\it 3.2. Development of tongues\hfil}
\medskip

In the above discussion we assumed that the equilibrium
meniscus shape (but not the height) at the center of the tube is established
at the end of stage (i). The rising in stage (ii) is
then governed by the unbalanced pressure drop, compensated
by the viscous forces in the liquid.
This description does not apply directly to the tongue region:
the equilibrium interface extends to infinite height which can not
be established in any finite time.
A more plausible picture is that, at a given time $t$,
there is a fully-developed tongue truncated at some height $h_m(t)$.
In addition, there can be an incipient tongue above $h_m(t)$.

A simple estimate for $h_m(t)$ can be made from Eq. (23).
The equilibrium thickness of the tongue at $h_m$ is
of the order of $a^2/h_m$ [see Eq. (6)]. Assuming that the viscous
flow up to this level is similar to the one in a tube of diameter
$D=a^2/h_m$, we obtain from Eq. (23)
$$h_m\simeq a(t/t_0)^{1/3}.\eqno(25)$$
Equivalently, the time for the tongue to reach a height $h$
is given by
$$\tau\simeq t_0(h/a)^3.\eqno(26)$$

It turns out that a more detailed calculation presented below
gives the same result as the simple-minded estimate (26).
Let us start with the approximate equation
$$(\partial^2_x+\partial^2_y)v_z=\phi(z),\eqno(27)$$
where
$$\phi(z)={1\over\nu}\bigl(g+{1\over\rho}\partial_z p\bigr).\eqno(28)$$
The term $\partial^2_z v_z$, which is omitted in (27), will be shown
to be small. In addition to the no-slip boundary condition $v_z=0$
on the solid walls, we demand that the tangential stress vanishes
on the liquid-gas interface, $\partial_n v_z=0$.

To fix the solution to (28), one has to specify the
shape of the cross-sectional area. For simplicity we assume that
all cross-sections of the tongue have the same shape,
and are parametrized only by the linear dimension $R(z)$.
In this case solution to (28) can be written as
$$v_z(x,y,z)=-\hat v(x/R,y/R)R^2\phi,\eqno(29)$$
where $\hat v$ is the solution to a dimensionless equation
which vanishes on the walls and is positive inside the region of interest.
We are now in a position to justify that $\partial^2_z v_z$ is
indeed negligible: it is smaller than $\phi$ by a factor $(R/z)^2$,
which is a small number in the tongue.

The final step is to write down an equation for $R(z,t)$, which can be
done using the continuity equation for an incompressible liquid,
$$\partial_t A+\partial_z U=0.\eqno(30)$$
Here $A=A_0 R^2$ is the cross-sectional area,
$U=\int dx dy\ v_z(x,y)=-U_0R^4\phi$ is the total velocity
in the vertical direction, and $A_0$ and $U_0$ are positive numbers which
depend only on the geometry.
Combining (30) with (28) and using $P=P_0-\gamma/R$, we obtain
$$\partial_t R^2=(U_0/A_0)\partial_z\bigl({g\over\nu}R^4+{\gamma\over\rho\nu}
R^2\partial_z R\bigr).\eqno(31)$$

Equation (31) has a stationary solution (6). We now consider a
linearized form of (31) around this solution. Writing
$R=(a^2/2z)(1+\delta)$, we obtain,
$$t_1\partial_t\delta={a^3\over z}\partial^2_z\delta-
{2a^3\over z^2}\partial_z\delta-{4a^3\over z^3}\delta,\eqno(32)$$
where $t_1=8t_0A_0/U_0$.
Equation (32) admits a scaling solution
$$\delta(z,t)=\Delta\bigl({(z/a)\over(t/t_1)^{1/3}}\bigr).\eqno(33)$$
The scaling function $\Delta$ satisfies
$${d^2\Delta\over du^2}+\bigl({1\over 3}u^2-{2\over u}\bigr){d\Delta\over du}
-{4\over u^2}\Delta=0.\eqno(34)$$
Thus the characteristic relaxation time for the build-up of the
fully-developed tongue at height $z$ is given by (26).

In the long-time limit, $\delta\rightarrow 0$, so that $\Delta(0)=0$.
The second boundary condition may be chosen to be
$\Delta(\infty)=-1$.
A numerical integration of (34) then yields the solution shown in
Fig. 4. For $u\gg 1$ we have $\Delta= -1+4u^{-3}+O(u^{-6})$.
In the opposite limit $u\ll 1$ we have
$\Delta\simeq -0.0317u^4[1-{1\over 18}u^3+O(u^6)]$.

\bigskip
\line{\bf 4. Discussion and Conclusions\hfil}
\medskip

In this paper we have shown that a liquid drop of finite extent
is unstable against spreading in a groove whose
opening angle is less than some critical value $\alpha_c$.
Spreading to infinite heights takes place even when the liquid has to climb
upwards against gravity. However, the thickness of the liquid tongue
decreases with increasing height. In addition, the total amount
of liquid in the tongue is finite: it is of the order of
$a^2D$ where $D$ is the diameter of the tube.

Since the equilibrium thickness of the tongues is proportional
to $a^2/z$ [Eq. (6)], at very high altitudes, microscopic
interactions such as van der Waals and double-layer forces become
important[1,3-5].
Taking $a=1$ mm and the range of molecular interactions
at $1 \mu$m, the macroscopic description used in this paper breaks down
when the height exceeds 1 m.
This regime is left for future investigation.

In the analysis presented on the dynamics of the rising process, we assumed
that the approach to equilibrium is controlled by viscous flow of the liquid,
which yields a $t^{1/2}$-law for rising of the center meniscus
and, at a later stage, a slower $t^{1/3}$-law for rising of the tongues.
These results presumably apply to very clean surfaces where contact
line hysteresis is negligible.
For pure water at room temperature, the surface tension and
kinematic viscocity coefficients are given by
$\gamma\simeq 73$ dyn/cm and $\nu\simeq 10^{-2}$cm$^2$/sec, respectively.
This yields $a=0.38$ cm and $t_0=\nu/(ga)\simeq 3\times 10^{-5}$ sec.
Using Eq. (26), we find that the time it takes for the liquid tongue to reach
1 cm, 10 cm, and 1 m are then $10^{-3}$ sec, 1 sec, and 10 min, respectively.
In our experiments with colored solutions, rising seems to be much
slower than that predicted above. The discrepancy at short times
is expected from the fact that the Reynolds number $R=vL/\nu$ is
quite high there. At later times, contact-line
pinning by dirts on the wall of the tube can influence the rising
process dramatically.
It is not clear, however, that the scaling laws should
break down when the velocity of the contact line is sufficiently high.
(For a discussion of contact line dynamics at low velocities
see Refs. [4,5,10].)

The analysis presented in this paper offers an explanation for
the fast release of bubbles in tubes with sharp grooves. Such tubes may
be more of a common occurrence than rarity in living bodies, for
which liquid transport is of crucial importance, and where capillary
forces are of relevance.
Our findings may also be of use in solving engineering problems
where trapping of bubbles is hazardous.
We hope our preliminary study will inspire further theoretical and
experimental investigations in this area, particularly on the dynamical side.

We wish to acknowledge useful discussions with
I. F. Lyusyutov and G. Nimtz. Upon completion of the work,
we discovered that wetting in a wedge near the liquid-gas transition
and in the absence of gravity
has been discussed previously, and the importance of the critical
groove angle $\alpha_c$ has been realized in that context[8,9].
We thank H. Dobbs for bringing these work to our attention.
The research is supported in part by the German Science
Foundation (DFG) under SFB 341.

\bigskip
\line{\bf References\hfil}
\medskip
\parskip=0truept
\item{[1]} See, e.g., Adamson A. W., {\it Physical Chemistry of Surfaces},
5th edition, John Wiley \& Sons, New York, 1990.

\item{[2]} Tang Y., Tang L.-H., Gu Y., Cao C., Wang B. and
Li S., submitted to Journal of Medical Biomechanics
(published by the Shanghai 2nd Medical University)(1993).

\item{[3]}Joanny J. F. and de Gennes P. G., J. Physique {\bf 47}, 121 (1986).

\item{[4]} de Gennes P. G., Rev. Mod. Phys. {\bf 57}, 827 (1985).

\item{[5]} Leger L. and Joanny J. F., Rep. Prog. Phys., 431 (1992).

\item{[6]} Landau L. D. and Lifshitz E. M., {\it Fluid Mechanics},
Pergamon, Oxford, 1959.

\item{[7]} In writing Eq. (17) we have assumed that the pressure
at $h=0$ inside the tube is the same as the atmospheric pressure.
This is not strictly true,
but the correction vanishes with vanishing flow rate.

\item{[8]} Pomeau Y., J. Colloid Interface Sci. {\bf 113}, 5 (1986).

\item{[9]} Hauge E. H., Phys. Rev. A {\bf 46}, 4994 (1992).

\item{[10]} Erta\c s D. and Kardar M., MIT preprint (1994);
Stepanow S., Tang L.-H. and Nattermann T., in preparation.
\bigskip
\line{\bf Figure Captions\hfil}
\medskip
\parindent=1.2cm
\item{Fig. 1.} Cross-sectional shape of tubes used in the experiment.
(b)-(e) contain sharp grooves along the tube.
The groove opening angles are: (b) $\alpha={1\over 2}\pi$;
(c) $\alpha=0$; (d) $\alpha={1\over 2}\pi$; (e) $\alpha={1\over 3}\pi$.

\item{Fig. 2.} (a) A wedge of angle $\alpha$ formed by two vertical
planes. Thin lines illustrate equilibrium capillary rise.
(b) Horizontal cross-section of (a). Liquid is confined in the
region $OCBDO$.

\item{Fig. 3.} Equilibrium profile of the tongue in an infinite wedge
of opening angle $\alpha={1\over 3}\pi$. The contact angle
is chosen at $\theta={1\over 12}\pi$.
(a) Horizontal cross-sections of the tongue
(thin solid lines) at heights $z/a=1.5$, 2, 2.5 and 3 (from right to left).
Dashed lines give the zeroth order approximation.
(b) The contact line (solid) on the wall and the vertical profile (dashed)
on the bisector of the wedge.

\item{Fig. 4.} The scaling function that appears in Eq. (33)
\vfil\eject
\bye